\documentclass[aps, prb, reprint, superscriptaddress]{revtex4-1}
\usepackage{graphicx}
\usepackage{dcolumn}
\usepackage{bm}
\usepackage{mathtools}

\begin{document}

\title{g-factor modification in a bulk InGaAs epilayer by an in-plane electric field}

\author{M. Luengo-Kovac}
\author{M. Macmahon}
\affiliation{ 
Department of Physics, University of Michigan
}

\author{S. Huang}
\affiliation{
Department of Materials Science and Engineering, University of Michigan
}

\author{R.S. Goldman}
\affiliation{
Department of Materials Science and Engineering, University of Michigan
}

\author{V. Sih}
\affiliation{
Department of Physics, University of Michigan
}

\date{\today}

\begin{abstract}
We report on the modification of the g-factor by an in-plane electric field in an In$_{0.031}$Ga$_{0.969}$As epilayer. We performed external magnetic field scans of the Kerr rotation of the InGaAs film in order to independently determine the g-factor and the spin-orbit fields. The g-factor increases from $-0.4473\pm0.0001$ at 0 V/cm to $-0.4419\pm0.0001$ at 25 V/cm applied along the \lbrack1$\overline{1}$0\rbrack\ crystal axis. In addition, spatially-resolved spin measurements show a g-factor dependence on diffusive velocity. The change in g-factor with electric field can have a large effect on the determination of the internal spin-orbit and nuclear fields from Larmor precession frequency measurements.
\end{abstract}

\maketitle

The field of spintronics is based on the manipulation of spins by electric and magnetic fields. The strength of the interaction of a spin with a magnetic field is determined by the g-factor. Therefore, local manipulation of the g-factor allows for local control of spins. Electron spin resonance (ESR) was shown to induce spin flips in a quantum dot when the frequency of an oscillating magnetic field matched the resonance frequency of the spin in a perpendicular static magnetic field, $g\mu_B B_{\text{stat}} = hf_{\text{osc}}$ \cite{Koppens_2006}. As this technique requires high magnitude, high frequency magnetic fields, alternative methods were developed that instead used an alternating electric field to produce the resonance, either by creating a oscillating spin-orbit (SO) field \cite{Schulte_2005} or by electrically modulating the g-tensor, called g-tensor modulation resonance (g-TMR) \cite{Kato_2003}. The use of an electric field also allows for more local manipulation than that possible with magnetic fields \cite{Nowack_2007}. 

g-TMR requires electrical control of the g-factor, which has been demonstrated in quantum wells (QWs) \cite{Salis_2001} and quantum dots (QDs) \cite{Deacon_2011} but not in bulk materials. In QWs and QDs, the change in the g-factor can be attributed to the shifting of the wave function into the barrier, which has a different g-factor, by an electric field. In this paper, we demonstrate electrical control of the g-factor in a bulk In$_{0.031}$Ga$_{0.969}$As epilayer. We characterize the g-factor dependence on the in-plane electric field and drift velocity using magnetic field and spatially-resolved pump-probe Kerr rotation spectroscopy. Our results for bulk InGaAs demonstrate that the electric field dependence here must be due to a different mechanism than the wave function shift and distinguish changes in the g-factor from changes in the spin-orbit field. A change in the g-factor with in-plane electric field was recently reported in a QW \cite{Chen_2014}.

Understanding the change in the g-factor as a function of the electric field is also important to correctly determine the magnitude of the internal effective magnetic fields produced by spin-orbit coupling and nuclear spin polarization. Fits for the internal fields from time-resolved Kerr rotation data \cite{Studer_2010, Trowbridge_2014, Meier_2007} assume that the g-factor is constant in order to separate it from the internal fields. In Ref. \onlinecite{Wilamowski_2007}, the g-factor is known not to change, but the precession frequency changes due to a change in the internal fields by an electric field. However, this is referred to as an effective tuning of the g-factor, using the relationship $g_{\text{eff}}(E) B_{\text{ext}} = g\left(B_{\text{ext}}+B_{\text{int}}(E)\right)$.

Measurements were performed on a 500 nm thick Si-doped In$_{0.031}$Ga$_{0.969}$As epilayer grown by molecular beam epitaxy on a (001) GaAs substrate. At room temperature, the carrier density was measured to be 1.45 $\times$ 10$^{16}$ cm$^{-3}$ and the mobility was 4870 cm$^2$/Vs.  Two channels 400 $\mu$m long and 100 $\mu$m wide were patterned and ohmic contacts were deposited to apply an in-plane voltage either along the [110] and [1$\overline{1}$0] directions. 

The g-factor of the spins is measured using a Kerr rotation setup. A mode-locked Titanium:Sapphire laser, with a 76 MHz repetition rate, tuned to 839.43 nm, is split into pump and probe beams. A mechanical delay line is used to control the temporal separation of the pump and probe. For our magnetic field-dependent measurements, the temporal separation of the pump and probe is set to 13 ns. In order to induce a spin polarization in the sample according to the optical selection rules \cite{Opt_Orient}, the pump is circularly polarized. The Kerr rotation of the linearly polarized probe is measured using a balanced photodiode bridge. The pump and probe are modulated by a photoelastic modulator and an optical chopper respectively for cascaded lock-in detection. The spatial separation of the pump and probe is controlled with a scanning mirror. 

For photoluminescence (PL) measurements, the sample is excited with a 633 nm HeNe laser. The spectrum is analyzed using a spectrometer with a silicon CCD. Both Kerr rotation and PL measurements are performed at 30 K, unless otherwise noted. 

Both time-resolved and magnetic field-dependent Kerr rotation measurements can be described by the equation \cite{Kikkawa_1998}:

\begin{equation} \label{RSAeq}
\theta_k(\Delta t, B_{ext}) = \sum_{n} A e^{-(\Delta t + nt_{\text{rep}})/T_2^*}\cos[\frac{g\mu_B}{\hbar} B_{\text{ext}}(\Delta t + nt_{\text{rep}})]
\end{equation}

\noindent where A is the Kerr rotation amplitude, $\Delta t$ is the time delay of the pump and probe, $t_{\text{rep}} = 13.16$ ns is the laser repetition rate, $T_2^*$ is the spin lifetime, g is the electron g-factor, $\mu_B$ is the Bohr magneton, $\hbar$ is the reduced Planck's constant, $B_{\text{ext}}$ is the magnitude of the external magnetic field, and n is the pulse number. 

When an in-plane voltage is applied across the sample, the spin packets created by subsequent pump pulses will be spatially separated due to carrier drift. Equation \ref{RSAeq} is modified to account for a spatially- and time-dependent amplitude $A_n(x, \Delta t)$. Furthermore, the spins experience a momentum-dependent spin-orbit splitting that acts like an effective internal magnetic field, requiring that the external magnetic field in Eq. \ref{RSAeq} be replaced by a vector sum of the internal and external magnetic fields \cite{Kato_2004}: 

\begin{multline}\label{SpinDrageq}
\theta_k(B_{ext},x) = \sum_{n} A_n(x, \Delta t) \times \\ 
\cos\left[\frac{g\mu_B}{\hbar} \sqrt{\left(\vec{B}_{\text{ext}}+\vec{B}_{\parallel}\right)^2+\vec{B}_{\perp}^2} (\Delta t + nt_{\text{rep}})\right]
\end{multline}

where $\vec{B}_{\parallel}$ and $\vec{B}_{\perp}$ are the components of the internal field parallel and perpendicular to the external field, respectively. 

Varying $\Delta t$ at a fixed $B_{\text{ext}}$ results in oscillations that are periodic in $\frac{g\mu_B}{\hbar}\sqrt{\left(\vec{B}_{\text{ext}}+\vec{B}_{\parallel}\right)^2+\vec{B}_{\perp}^2}$. If instead $\Delta t$ is fixed and $B_{\text{ext}}$ is varied, then the oscillations are periodic in $\frac{g\mu_B}{\hbar}\Delta t$.

\begin{figure}
\includegraphics[width=8.5cm]{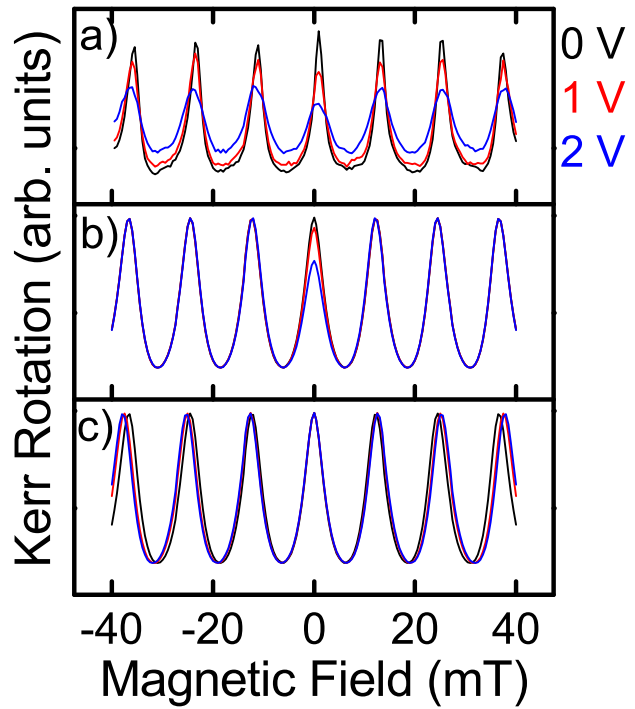}
\caption{\label{fig1} (a) Kerr rotation data as a function of the external magnetic field for applied voltages of 0 V (black), 1 V (red), and 2 V (blue) V fit to Eq. 2. The fit values were then plugged back into Eq. 2 and plotted varying only the spin-orbit field (b) and only the g-factor (c) as a function of voltage to show the separate dependences on these two variables. }
\end{figure}

The components of the spin-orbit field parallel and perpendicular to the external field shift the curve and decrease the magnitude of the center peak respectively (Fig. \ref{fig1}a,b), whereas the g-factor changes the frequency of the peaks (Fig. \ref{fig1}c). In this way, unlike in time-resolved measurements, we can separately determine the g-factor and the spin-orbit field. 

\begin{figure}
\includegraphics[width=8.5cm]{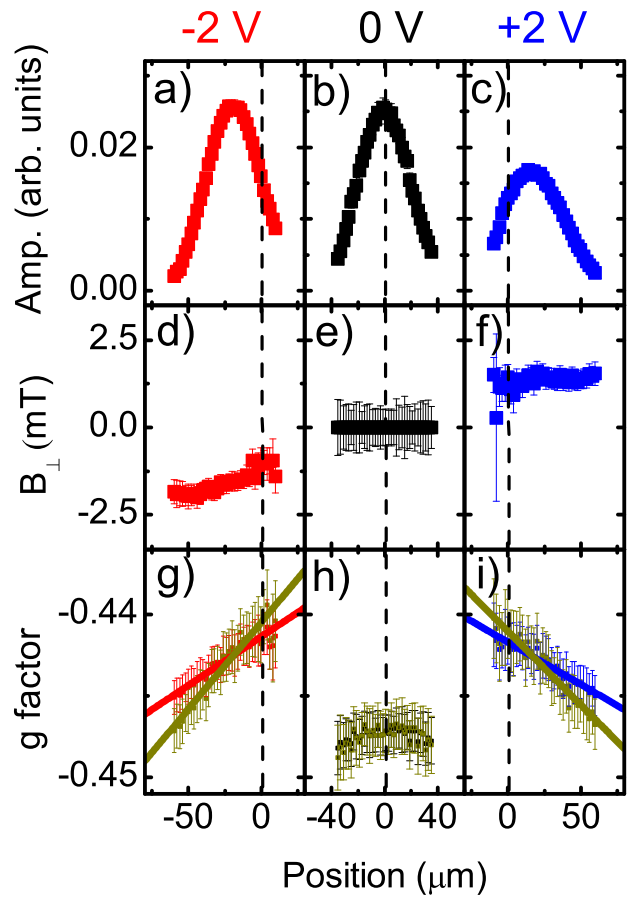}
\caption{\label{fig2}(a), (b), (c) Amplitude of the Kerr rotation, (d), (e), (f) fits of the perpendicular component of the internal field, and (g), (h), (i) fits of the g-factor as a function of pump-probe separation for -2 V, 0 V, and +2 V.  The spatial dependence of the g-factor that indicates a diffusive velocity dependence. Corrected g-factor values found using Eq. 4 are shown in gold. }
\end{figure}

Magnetic field scans were performed at different pump-probe separations, with the external magnetic field parallel to the applied voltage. The amplitude as a function of this separation (Fig. \ref{fig2}a,b,c) shows the profile of the spin packet. The difference in amplitude between the +2 V and the -2 V scans is due to drift of various factors over the course of the measurements, including laser power, vertical alignment of the pump, and the balance of the photodiode bridge. However, the absolute amplitude doesn't affect the subsequent fits. 

The average drift velocity ($v_d$) of the electrons can be calculated as a function of voltage from the center position ($x_c$) of the amplitude and the known temporal separation of the pump and probe pulses. From measurements of the drift velocity as a function of applied electric field, we determine the effective mobility, using the equation $\mu_{\text{eff}} = v_d/E_{\text{applied}}$, to be $5110 \pm 70$ cm$^2$/(V s) along the [110] direction and $5800 \pm 200$ cm$^2$/(V s) along the [1$\overline{1}$0] direction. The difference in mobilities along the two directions comes from a discrepancy between the total applied voltage and the actual voltage across the channel, as there is likely also a voltage drop across the contacts that is different for different contacts. As such, all electric field dependent measurements are presented in terms of the drift velocity, which we can measure directly. 

SO field strength measurements were performed as in Ref. [\onlinecite{Norman_2010}] (Fig. 2d,e,f). The internal fields were found to be perpendicular to the direction of the current for both crystal axes, as is expected for these directions. The spin-orbit field proportionality constants were found to be 1.96 $\pm$ 0.02 mT ns $\mu$m$^{-1}$ along [110] and 1.02 $\pm$ 0.02 mT ns $\mu$m$^{-1}$ along [1$\overline{1}$0] (Fig. 3b). 

The g-factor was extracted from the external magnetic field scan at each pump-probe spatial separation (Fig. \ref{fig2}g,h,i). The position dependence of the g-factor is likely related to the velocity-dependent spatial separation of the spins, indicating that there is some diffusive velocity dependence to the g-factor. 

Since the probe beam has some width $w_p$, the measured g-factor at a given pump-probe separation is an average of the g-factor of the electrons within the probe spot. Given the probe and spin packet widths, we can deconvolute the effects of the finite probe size from the measured g-factor to get the corrected g-factor. Since the probe beam and spin packet both have a Gaussian profile, we can write the measured g-factor as a function of the corrected g-factor \cite{Norman_Thesis}:

\begin{equation} \label{meas_g}
g_{\text{meas}}(x_p) = \frac{\int_{-\infty}^{\infty}{g(x)e^{-\frac{(x-x_d)^2}{2w_d^2}} e^{-\frac{(x-x_p)^2}{2w_p^2}} } \mathrm{d} x } {\int_{-\infty}^{\infty}{e^{-\frac{(x-x_d)^2}{2w_d^2}} e^{-\frac{(x-x_p)^2}{2w_p^2}} }\mathrm{d} x}
\end{equation}

where $w_d$ ($w_p$) is the width of the spin packet (probe) and $x_d$ ($x_p$) is the position of the spin packet (probe). Since the measured g-factor has a linear dependence on position, we can assume that the corrected g-factor has the same dependence. Using this assumption, we can integrate Eq. \ref{meas_g} and rearrange to solve for the corrected g-factor:

\begin{equation} \label{act_g}
g(x) = g_{\text{meas}}(x) + \left(\frac{w_p}{w_d}\right)^2(g_{\text{meas}}(x)-g_{\text{meas}}(x_d))
\end{equation}

Both the measured g-factor and the corrected g-factor, calculated in this way, are shown in Fig. 2g,h,i, using $w_p = 17.5 \pm 0.5$ $\mu$m and $w_d$ determined from a deconvolution of the probe with the amplitude profiles in Fig. 1a,b,c. One can see that the deconvolution of the g-factor increases the change in the g-factor as a function of position. Since the pump is at x = 0 $\mu \text{m}$, spins that are farther from the pump location have a higher diffusive velocity. Thus, we can conclude that faster moving electrons have a g-factor that is larger in magnitude.

\begin{figure}
\includegraphics[width=8.5cm]{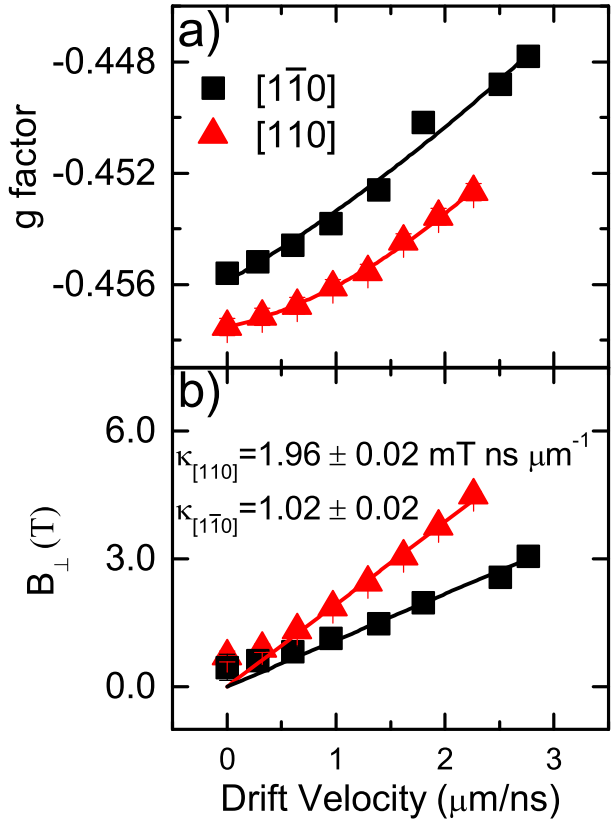}
\caption{\label{fig3}(a) g-factor as a function of electric field (shown in terms of the drift velocity) for the [110] and [1-10] directions. The parabolic fits to the g-factor are a guide to the eye and are less accurate a higher drift velocities. The offset is likely due to g-factor anisotropy. (b) The internal field magnitude as a function of drift velocity. The slopes of the lines give the spin-orbit field proportionality constant.  }
\end{figure}

The g-factor at the center of the spin packet is plotted versus the drift velocity $v_d$ (Fig. \ref{fig3}a). Parabolic fits are shown as a guide to the eye. However, a numerical derivative shows that this fit is less accurate for larger drift velocities. Measurements conducted along the [110] and [1$\overline{1}$0] crystal axes show similar curvatures:  (1.04 $\pm$ 0.7) $\times 10^{-3}$/($\frac{\mu\text{m}}{\text{ns}})^2$ for [110] and (1.23 $\pm$ 0.2) $\times 10^{-3}$/($\frac{\mu\text{m}}{\text{ns}})^2$ for [1$\overline{1}$0]. On the other hand, the spin-orbit field proportionality constant is about twice as large along the [110] direction. The offset is likely due to the anisotropic g-factor tensor \cite{Kato_2003}. 

It is interesting to note that we see two opposite effects on the g-factor due to diffusive velocity and the applied electric field. Figures \ref{fig2}g,i show that spins that have experienced a larger net displacement have a higher magnitude g-factor, whereas Fig. \ref{fig3}a shows that electrons in a larger electric field have a smaller magnitude g-factor. 

Ref. \onlinecite{Hubner_2009} indicates that the energy dependence of the g-factor in GaAs is given by $g^{*}(E) = g^{*} + 6.3 \text{ eV}^{-1}\cdot E$. If we assume a parabolic dispersion relation for our values of $k$, then our measured dependence on the energy, taken from electric-field dependent measurements, is more than 500 times larger. Furthermore, we see the opposite trend for electrons of greater speed. 

\begin{figure}
\includegraphics[width=8.5cm]{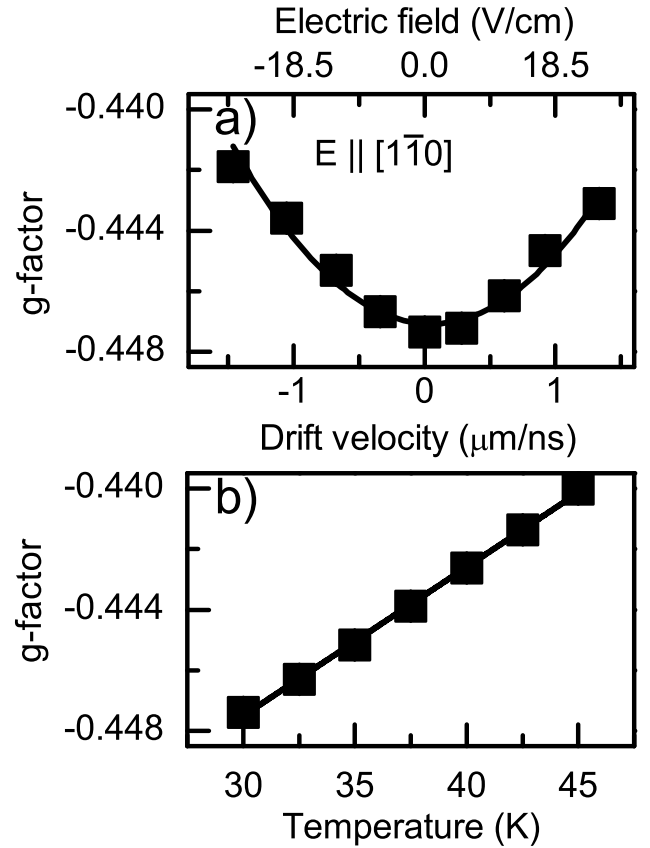}
\caption{\label{fig4}(a) g-factor as a function of drift velocity along [1-10] at 30K, with a parabolic fit as a guide to the eye. (b) Temperature dependence of the g-factor for vd = 0. The change in g-factor due to an applied voltage of 1 V at 30 K corresponds to the change to the channel heating by 9 K.  }
\end{figure}

Measurements for both positive and negative $v_d$ shows that the change is g-factor is symmetric about zero drift velocity (Fig. \ref{fig4}a), and thus only depends on the magnitude of $v_d$, not the direction. Temperature-dependent measurements of the g-factor (Fig. \ref{fig4}b) were performed on the same channel in order to compare the effects of temperature and electric field. The change in g-factor due to an applied voltage of 1 V at 30 K is equivalent to the change due to the channel heating by 9 K. From power dissipation calculations, we can estimate that the expected change in temperature of the channel due to an applied voltage of 1 V is $\mathcal{O} (10^{-4})$ K. Therefore, the g-factor dependence on voltage and temperature are likely distinct phenomena. 

To check whether the applied voltage was causing excessive channel heating, we performed temperature and voltage dependent PL measurements. From 30 K to 40 K, there was a clear shift in the energy of the peak due to the change in the bandgap with lattice temperature. However, the PL for 0 V and 2 V had no discernible shift. Therefore, it is unlikely that the applied voltage is causing sufficient channel heating to account for the change in the g-factor. 

We expect the electron temperature to be significantly above the lattice temperature for our range of applied electric fields, and one possible explanation for the change in g-factor is that the g-factor is dependent on the electron temperature. It was reported that in n-GaAs the electron temperature increased from 4.2 K at 0 V/cm to 38 K at 20 V/cm and 75 K at 50 V/cm \cite{Aoki_1978}. The electron temperature was estimated from the high energy tail of the PL. We saw no change in the PL of our sample as a function of electric field. However, we expect the PL of our sample to have greater broadening due to it being an alloy. The broadening due to alloy fluctuations could be overwhelming any changes due to the electric field. 

The electron temperature can be estimated if the energy loss rate per electron is known, based on measurements done on AlGaAs/GaAs heterostructures using PL \cite{Shah_1983}, Shubnikov-de Haas Oscillations \cite{Sakaki_1984}, and far-infrared spectroscopy \cite{Hirakawa_1993}. For an applied voltage of 2 V, we found the energy loss rate per electron in our samples to be $2\times10^{-12}$ W, which corresponds to an electron temperature of about 50 K for a lattice temperature of 4.2 K. 

Another possible explanation is that the applied electric field is modifying the wave-function of donor-bound electrons, causing their wave functions to spread farther into the semiconductor, thus changing the g-factor \cite{De_2009}. Calculations for a Si dopant in GaAs show that the relative magnitude change in the g-factor is comparable to what we measure here. However, the sign of the change is opposite. Furthermore, for the doping density of the sample and at the temperatures considered here, contributions from donor-bound electrons are expected to be minimal. Therefore we can rule out modification of donor-bound electron wave functions as the cause of the g-factor modification. 

The orbital contribution of spin-orbit induced circulating currents was shown to be significant in calculations of the g-factor in quantum dots \cite{vanBree_2014}. Similarly, the net drift velocity of the spins in our sample could be modifying the orbital contribution to the g-factor. 

We have performed electric-field dependent measurements of the g-factor in a bulk In$_{0.031}$Ga$_{0.969}$As epilayer in a manner that distinguishes between changes in the g-factor from changes in the spin-orbit field. Separate determination of these two quantities is important as their percent change with voltage is comparable. For example, for measurements along the [110] direction (Fig. 3), we found the g-factor to be $-0.45268\pm0.0003$ and the SO-field to be $4.50\pm0.08$ mT for the largest drift velocity. If instead time-resolved measurements had been done to determine the SO field from the Larmor precession frequency with an applied magnetic field of 0.2 T assuming the zero field value of the g-factor, the SO field would have been calculated to be only $2.33\pm0.20$ mT. However, more work is needed to develop a quantitative model for this phenomenon. 

This work was supported by the National Science Foundation Materials Research Science and Engineering Center program DMR-1120923, the Office of Naval Research, the Air Force Office of Scientific Research, and the Defense Threat Reduction Agency Basic Research Award No. HDTRA1-13-1-0013. SH and RSG were supported in part by the Center for Solar and Thermal Energy Conversion, an Energy Frontier Research Center funded by the U.S. Department of Energy Office of Science, Office of Basic Energy Sciences under Award No. DE-SC0000957.

\newpage 

\end{document}